\documentclass[prl,a4paper,superscriptaddress,twocolumn]{revtex4}
\usepackage{color,amsmath,amssymb,amsthm,times,graphics,graphicx,bm,dcolumn}

\newcommand{\be}{\begin{equation}}
\newcommand{\ee}{\end{equation}}
\newcommand{\ben}{\begin{eqnarray}}
\newcommand{\een}{\end{eqnarray}}
\newcommand{\bes}{\begin{subequations}}
\newcommand{\ees}{\end{subequations}}

\begin{document}
\title{Efficient simulation of strong system-environment interactions}

\author{Javier Prior}
  \affiliation{Departamento de F\'{\i}sica Aplicada, Universidad Polit\'ecnica de Cartagena, Cartagena 30202, Spain}
\affiliation{QOLS, The Blackett Laboratory, Prince Consort Road, Imperial College, London, SW7 2BW, UK}

\author{Alex W. Chin}
 \affiliation{Institut f\"{u}r Theoretische Physik, Albert-Einstein-Allee 11, Universit\"{a}t Ulm, D-89069 Ulm, Germany}

\author{Susana F. Huelga}
 \affiliation{Institut f\"{u}r Theoretische Physik, Albert-Einstein-Allee 11, Universit\"{a}t Ulm, D-89069 Ulm, Germany}

\author{Martin B. Plenio}
  \affiliation{Institut f\"{u}r Theoretische Physik, Albert-Einstein-Allee 11, Universit\"{a}t Ulm, D-89069 Ulm, Germany}
\affiliation{QOLS, The Blackett Laboratory, Prince Consort Road, Imperial College, London, SW7 2BW, UK}

\begin{abstract}
Multi-component quantum systems in strong interaction with their environment
are receiving increasing attention due to their importance in a variety of contexts, ranging from solid state quantum information processing to the
quantum dynamics of bio-molecular aggregates. Unfortunately, these systems are difficult to simulate as the system-bath interactions cannot be treated perturbatively and standard approaches are invalid or inefficient.
Here we combine the time dependent density matrix renormalization
group methods with techniques from the theory of orthogonal polynomials to
provide an efficient method for simulating open quantum systems, including spin-boson models and their generalisations to multi-component systems. 
\end{abstract}



\maketitle

{\em Introduction--}
Dissipation and decoherence caused by noisy, fluctuating environments
are fundamental and ubiquitous phenomena which appear in almost all experiments on quantum systems.
While initially thought to lead to the rapid erasure of quantum signatures,
it was later realized that the presence of environmental noise can in
fact be instrumental for maintaining quantum correlations in the steady
state \cite{prl02}. Decoherence is also thought to be essential for
explaining the efficient excitation energy transport (EET) in the
pigment-protein complexes (PPCs) of photosynthetic organisms \cite{aspuruguzik08a,plenio08}, biological systems in which long-lived quantum coherence has been recently observed \cite{brixner05}.
As a result, novel emphasis is being placed in better understanding the
interplay between the coherent quantum dynamics and incoherent processes
arising from the interaction with the environment \cite{nat}. Ultimately,
this understanding may facilitate the development of noise-assisted
devices and, potentially, light-harvesting systems with increased efficiency.

To date, the dynamical behaviour of interacting open quantum systems and,
in particular, noise-assisted transport, is frequently investigated
in terms of simple dynamical models in which environmental
dephasing and relaxation are treated with Lindblad or Bloch-Redfield
master equations. These standard methods are both based on the assumptions
of weak system-bath coupling and the Markov approximation. However,
these approximations are not valid for many realistic systems of current
interest, and assuming that the correlation time of the environments in these examples is much faster
than the system dynamics is frequently not justified. For instance, in typical
PPCs the dynamical timescales of the bath can be comparable or even slower
than
the EET dynamics \cite{ishizaki09,thorwart09}. Moreover, in the limit
of slow bath dynamics, perturbative treatments of the system-environment
coupling cannot be used even if the system-bath coupling is intrinsically
weak \cite{ishizaki09,nalbach09}. Recently, important steps have been taken towards developing
and analysing fully non-perturbative and non-markovian approaches, including the numerical renormalisation group (NRG) \cite{anders07},
the numerical hierarchy technique (NHT) \cite{ishizaki09}, and the numerical path integral (NPI) \cite{thorwart09}. There are, however, limitations concerning the
allowed spectral density of the bath and these techniques are expected to become less efficient with decreasing temperatures and highly structured environments. \\
Given the general lack of information about the real protein spectral
densities in PPCs, a technique is required that can simulate EET for
arbitrary spectral densities and coupling strengths, thus allowing experiments carried out under different conditions, including low temperatures,
to be analyzed within one framework. A similar situation is encountered in solid state
qubit implementations, where the microscopic details behind the dephasing caused by spurious two level fluctuators
still needs to be fully clarified \cite{kit}.\\
Here we address these issues by developing a numerically exact and
efficient method that combines time-adaptive density matrix renormalisation
(t-DMRG) with analytical transformation techniques based on the theory of
orthogonal polynomials. This method delivers both
controllable accuracy and numerical efficiency, without any restrictions
on the complexity or strength of the system-environment coupling. Even though
we will present numerical examples of richly structured environments used in the PPC literature, it should be emphasized that this new simulation 
tool is completely general, and can be applied to any system linearly coupled to bosonic or fermionic environments. Our method also provides
complete information about the evolving state of the environment, and opens
the door to detailed studies of the system-bath correlations which give rise
to long-lasting coherences, entanglement and other novel effects. 

{\em The Model--}
We demonstrate our approach to open-system dynamics by considering an
elementary model of a PPC, a dimer molecule consisting of two pigments
(henceforth referred to as `sites'). The dimer's internal dynamics are described by a Hamiltonian $H_{S}=\frac{\epsilon_{1}}{2}\sigma_{1z}+\frac{\epsilon_{2}}{2}\sigma_{2z}+J(\sigma_{1+}\sigma_{2-}+\sigma_{2+}\sigma_{1-})$, where $\sigma_{i+},\sigma_{i-},\sigma_{iz}$ are standard Pauli
creation, annihilation, and $z$ matrices for the $i$th site of the dimer.
The spin-down state $\sigma_{iz}|\downarrow_{i}\rangle=-|\downarrow_{i}\rangle$
represents the ground state of the site and the spin-up state
represents a single local excitation which can hop between sites with a tunneling amplitude $J$. These sites interact with local
independent environments modeled as continuous baths of harmonic oscillators,
and the site-environment interaction $H_{I}$ and the environment Hamitlonian $H_{B}$ can be written as
\begin{eqnarray}
        H_{I}&=&\sum_{i=1,2}\frac{(1+\sigma_{iz})}{4}\int_{0}^{1}h(k)(a_{i}(k)
        +a_{i}^{\dagger}(k))dk,\label{HI}\\
        H_{B}&=&\sum_{i=1,2}\int_{0}^{1}g(k)a_{i}^{\dagger}(k)a_{i}(k) dk, \label{HB}
\end{eqnarray}
where the environmental modes for each site $i$ are described by creation and annihilation
operators $a_{i}^{\dagger}(k),a_{i}(k)$ respectively, and satisfy the
continuum commutation relation $[a_{j}(k),a_{j}^{\dagger}(k')]=\delta_{ij}
\delta(k-k')$. For simplicity we assume that both baths have identical dispersions $g(k)$
and coupling strengths $h(k)$, though our method could easily accommodate different coupling structures on
each site. We further assume that the spectrum of the bath frequencies is limited by a high-frequency cut-off $\omega_{c}$. The form of the
coupling is chosen so that it is zero when a site is in the ground state.
\begin{figure}
\includegraphics[width=8.5cm]{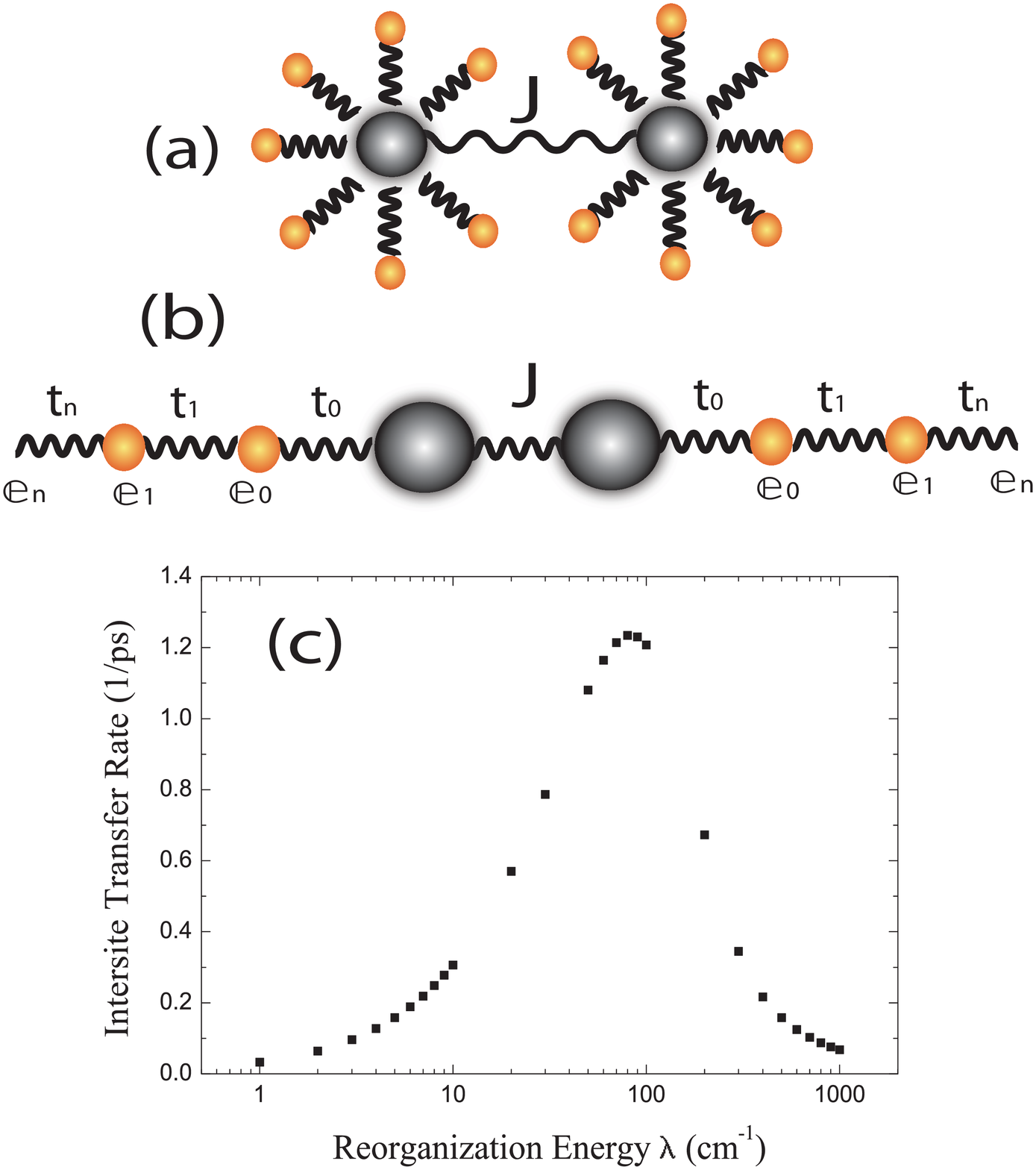}
\caption{(a) The standard Hamiltonian has each site of the dimer interacting with its surrounding bath in a star-like configuration. (b) After a unitary transformation of the bosonic modes only, an infinite harmonic $1-D$ chain Hamiltonian is generated with frequencies $\epsilon_{n}$ and nearest-neighbour couplings $t_{n}$. The dynamics of the dimer is the same in both (a) and (b), but the Hamiltonian of (b) can be efficiently simulated using t-DMRG. (c) Intersite population transfer rate $\Gamma$ as a function of reorganisation energy $\lambda$ for the spectral density of Eq. (\ref{jobo}) at $T=0$ K. Simulation parameters are $J=20\textrm{cm}^{-1},\epsilon_{1}-\epsilon_{2}= 100\textrm{cm}^{-1}$ and $\gamma=53 \textrm{cm}^{-1}$. Due to the presence of multiple timescales in the population transfer, $\Gamma$ was estimated as the average decay rate of the populations at time $1.3$ ps. }
\label{fig1}
\end{figure}
The effects of the site-bath interactions are completely determined by
specifying the spectral function $J(\omega)$ of the bath \cite{leggett}.
Introducing the inverse function for the dispersion $g^{-1}(k)$, $J(\omega)$
is defined for the continuous oscillators by $J(\omega)=\pi h^{2}(g^{-1}(\omega))
\frac{dg^{-1}(\omega)}{d\omega} $ \cite{bulla}. A given $J(\omega)$ does not
uniquely specify both $g(k)$ and $h(k)$, but we now show how the
choice $g(k)=\omega_{c}k$ (corresponding to a linear sampling of the spectral
density) allows the original Hamiltonian to be mapped into a $1D$ chain
form suitable for efficient DMRG simulation. Note that the specific form of
the system operator in Eq. (\ref{HI}) is of no importance for the following
transformations, and the mapping will also apply, for example,
for environments that induce or suppress spontaneous decay of excitations such as those found in photonic crystals \cite{angelakis04}. 

We can represent the continuum operators by a new set of bosonic operators
labelled by a discrete index. These
new bosonic operators are given by  $b_{in}=\int_{0}^{1}U_{n}(k)a_{i}(k) dk$
and $b_{in}^{\dagger}=\int_{0}^{1}U_{n}(k)a_{i}^{\dagger}(k) dk$, where
$n=0,1,2...\infty$. We now choose $U_{n}(k)=h(k)\rho_{n}^{-1}\pi_{n}(k)$, where $\rho_{n}$ is a normalisation constant, and
$\pi_{n}(k)$ is the $n$th monic polynomial of a sequence of orthogonal polynomials defined
by the inner product $\int_{0}^{1}h^{2}(k)\pi_{n}(k)\pi_{m}(k)dk=\rho_{n}^{2}
\delta_{nm}$. With the choice $g(k)=\omega_{c}k$, we have $h^{2}(k)= \pi^{-1}
\omega_{c}J(g(k))$. Using the properties of orthogonal
polynomials it is easy to show that the transformation must be
orthogonal and that the new modes obey bosonic commutation relation
$[b_{n},b_{m}^{\dagger}]=\delta_{nm}$ \cite{chin10}. Expressing $H$ in
terms of the new modes, a $1D$ chain Hamiltonian of the form shown in
Fig.\ref{fig1} is generated, and is given by $\tilde{H}=H_{S}+H_{IC}+H_{C}$,
\begin{eqnarray}
H_{CI}&=&\sum_{i=1,2}\sqrt{\frac{\eta}{\pi}}\frac{(1+\sigma_{iz})}{4}(b_{i0}+b_{i0}^{\dagger}),\\
H_{C}&=&\sum_{i=1,2}\sum_{n=0}^{\infty}\epsilon_{in}b_{in}^{\dagger}b_{in}+(t_{in}b_{in+1}^{\dagger}b_{in}+ h.c.)\label{hc}.
\end{eqnarray}
The new modes of the chain have frequencies $\epsilon_{jn}$ and nearest-neighbour couplings $t_{jn}$. The coupling of site $j$ to the first member of the chain
is determined by $\eta=\int_{0}^{\omega_{c}}J(\omega)\,d\omega$. The
nearest-neighbour chain structure is a direct consequence of using the basis of
orthogonal polynomials. When the new operators are substituted into Eq.(\ref{HI}) the effective coupling
strength of the $nth$ mode of the chain to the site is $\propto \int_{0}^{1}h^{2}(k)
\pi_{n}(k)\,dk$, which is only non-zero for $n=0$ due to the orthogonality of the $\pi_{n}(x)$ and the fact that $\pi_{0}(x)=1$. Similarly, when the new modes are substituted into
Eq.(\ref{HB}), modes $n$ and $m$ couple to each other with a strength $\propto
\int_{0}^{1}h^{2}(k)\,k\,\pi_{n}(k)\pi_{m}(k)\,dk=\int_{0}^{1}h^{2}(k)\,
(\alpha_{n}\pi_{n}(k)+\beta_{n}\pi_{n-1}(k)+\pi_{n+1}(k))\pi_{m}(k)\,dk$,
where we have used the standard three-term recurrence relation obeyed by
all monic orthogonal polynomials to replace the product $k\pi_{n}(k)$. After this step,
orthogonality guaranties nearest-neighbour couplings only, and the whole system of sites and oscillators can be arranged in an
infinite $1D$ chain as shown in Fig.\ref{fig1}. Note that we have chosen
a dimer setting only for simplicity, and that systems with more components can also be treated efficiently using known generalizations of DMRG.

This chain mapping is exact and does not require any discretisation of the spectral density before being applied.The transformation, which also works for fermions, and its analytical properties will presented rigorously in Ref. \cite{chin10}, where analytical formulae for $\epsilon_{n},t_{n}$ can be found for both the linear DMRG and the logarithimically discretised NRG chain mappings for $J(\omega)\propto \omega^{s}$. \cite{bulla}. When analytical results are not available, the use of orthogonal polynomials also offers significant numerical advantages, as it can be shown that $\epsilon_{in}$ and $t_{in}$ are simply related to the recurrence coefficients $\alpha_{n},\beta_{n}$ introduced above, and can be determined accurately using very stable numerical algorithms such as the ORTHOPOL package \cite{chin10,gautschi}. Interestingly, mathematical theorems on these transformations demonstrate that for \textit{any} smooth spectral density which is strictly bigger than zero in its domain with cut-off $\omega_{c}$, $\epsilon_{in} \rightarrow\omega_{c}/2,
t_{in}\rightarrow\omega_{c}/4$ as $n\rightarrow\infty$. This result is the mathematical expression of the fact that for a smoth spectral density we expect that excitations are eventually lost to the environment irreversibly. In the linear chain picture this irreversibility arises because a translationally invariant harmonic chain with
this ratio of $\epsilon_n$ to $t_n$ has a gapless spectrum and can carry away arbitrary excitations with energies up to $\omega_{c}$. Thus there
is no unlimited build-up of excitations in the chain and correlations
are expected to be bounded. This in turn suggests that DMRG will converge
quickly \cite{EisertCP10}.

{\em The overdamped brownian oscillator spectral density --}
We now consider some specific spectral densities of relevance for PPCs in
photosynthetic organisms. To start with, we look at the overdamped Brownian
oscillator spectral density which has previously been studied by Ishizaki
and Fleming in the high temperature limit \cite{ishizaki09}.
The bath in our simulations is at $T=0$ K for simplicity, although we can
extend our method to finite temperatures using standard extensions of
DMRG to mixed states \cite{rosenbach}. In our notation,
the overdamped brownian oscillator spectral density has a simple Ohmic form,
\begin{equation}
J(\omega)=\frac{8\lambda\gamma\omega}{\omega^{2}+\gamma^2},
\label{jobo}
\end{equation}
where $\lambda$ is the reorganisation energy of the bath, defined by $\lambda=\frac{1}{4\pi}\int_{0}^{\omega_{c}}J(\omega)\omega^{-1}d\omega$, and is taken as our measure of the site-environment coupling strength. The parameter $\gamma$ approximately sets the dynamical response time of the bath, and the following simulations use values of $\gamma$ which are smaller than the dimer energy scales in order to observe non-markovian effects \cite{ishizaki09}. In a PPC an initial excitation is injected suddenly onto a site from either another PPC or the capture of a photon. The initial condition of the subsequent dynamics at $T=0$ K is the separable state of zero excitations in the bath, and one excitation on site one. This initial condition is used throughout this work, although our approach allows us to simulate the evolution of any initial state, including highly correlated states of the system and bath.

Figure (\ref{fig1})c shows the average transfer rate of the population initially on site $1$ to site $2$ as a function of $\lambda$. We see here that the transfer rate is a non-monotonic function of $\lambda$, with an optimum transfer rate occurring for $\lambda\approx 80 \mathrm{cm}^{-1}$. Ishizaki and Fleming studied the same dimer and bath parameters at $300$ K using their NHT and found completely incoherent and exponential relaxation for $\lambda>5 \mathrm{cm}^{-1}$, with the resulting intersite rate also being a non-monotonic function of $\lambda$ with a maximum at $\lambda\approx 10-20 \mathrm{cm}^{-1}$. This non-monotonicity is a signature of the non-perturbative nature of both simulations, and in the cases of the NHT, it shows that the method interpolates correctly between a weak-coupling Redfield-like theory and the incoherent F\"{o}rster theory where multi-phonon effects dominate \cite{ishizaki09}. However, the DMRG population evolutions (not shown) show that coherence is more robust at $T=0$ K, with weakly-damped oscillations for $\lambda\leq 30 \mathrm{cm}^{-1}$, followed by incoherent relaxation with multiple decay rates for $30\geq\lambda\leq 50 \mathrm{cm}^{-1}$. For larger $\lambda$ the localisation effect discussed below comes increasingly into play. 

\begin{figure}
\includegraphics[width=8.5cm]{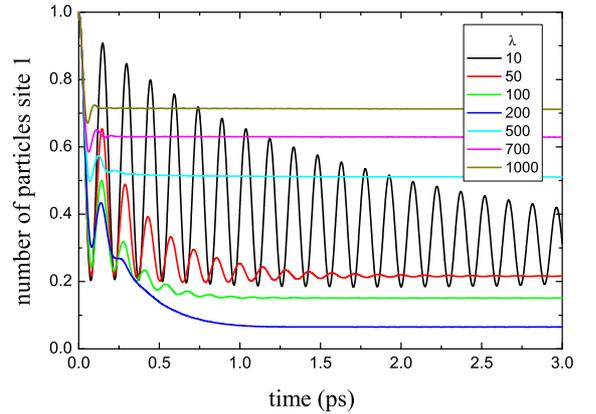}
\caption{Evolutions of the population on site $1$ for the spectral density of Eq. (\ref{jobo}) at $T=0$ K., and various reorganisation energies $\lambda$. Simulation parameters are $J=100\textrm{cm}^{-1},\epsilon_{1}-\epsilon_{2}= 100\textrm{cm}^{-1}$ and $\gamma=53 \textrm{cm}^{-1}$}
\label{fig3}
\end{figure}

Figure (\ref{fig3}) shows the population on site 1 as a function of time for various $\lambda$s. For $\lambda\leq 100
\mathrm{cm}^{-1}$ we find damped oscillations which persist for at least $1$ ps.
For larger $\lambda$, coherent dynamics are always seen for a few hundred femtoseconds
before the dynamics becomes incoherent, although as $\lambda$ increases the duration of coherent motion becomes shorter. For $\lambda\geq 200
\mathrm{cm}^{-1}$ the incoherent relaxation rate decreases dramatically, and
an increasingly large population is trapped on site $1$
over the timescale of the simulations. This quantum-zeno-like phenomenon may be related to the well-studied localisation transition
found in Ohmic and sub-Ohmic spin-boson models at $T=0$ K \cite{leggett}. This is another
non-perturbative feature of the dynamics, and similar dynamics have also recently been
observed in NRG and NPI studies of the sub-Ohmic spin-boson model \cite{anders07,thorwart10}.

{\em Other spectral densities --}
We now demonstrate the versatility of our method w.r.t. the microscopic system-bath interactions by considering a much more complex and structured environmental spectral function taken from a recent study of photosynthetic EET. In Ref \cite{adolphs06}, Adolphs and Renger use a combination of super-ohmic densities and a coupling to a single effective high-energy mode to model the environment. In our notation this spectral function can be written as,
\begin{eqnarray}
J(\omega)&=&\frac{2\pi\lambda\,[1000\omega^{5}e^{-\left(\frac{\omega}{\omega_{1}}\right)^{\frac{1}{2}}}+4.3\omega^{5}e^{-\left(\frac{\omega}{\omega_{2}}\right)^{\frac{1}{2}}}]}{9!(1000\omega_{1}^{5}+4.3\omega_{2}^{5})}\nonumber\\
&+&4\pi S_{H}\omega_{H}^{2}\delta(\omega-\omega_{H})\label{JAR},
\end{eqnarray}
where we have kept the relative contributions of the two continuous parts of the spectral density as they are in \cite{adolphs06}, but have also introduced an overall reorganisation energy $\lambda$ to be used as a free parameter. The coupling to the high-energy mode is fixed, and the parameters of the simulation are $J=100\textrm{cm}^{-1},\epsilon_{1}-\epsilon_{2}= 100\textrm{cm}^{-1},\omega_{1}=0.5 \textrm{cm}^{-1}, \omega_{2}=1.95 \textrm{cm}^{-1}, \omega_{H}=180 \textrm{cm}^{-1}, \omega_{c}=1000 \textrm{cm}^{-1}$ and $S_{H}= 0.22$ \cite{adolphs06}. With these values the continuous part of $J(\omega)$ extends over a frequency range of about $900 \mathrm{cm}^{-1}$, and $\omega_{H}$ is almost resonant with the energy difference ($ ~224\mathrm{cm}^{-1} $) of the dimer eigenstates of $H_{s}$ as the coupling strength of this mode to a site is $~84 \mathrm{cm}^{-1}$.  

The chain transformation and DMRG method offers numerical advantages over some other techniques for spectral functions which contain delta functions or damped resonances, as strong coupling to such modes of the environment do not have to be considered part of the quantum system itself. A simple modification of the chain parameters allows simulation of an arbitrary number of such mode interactions with no increase in the complexity of the simulation. Coupling to undamped modes with frequencies comparable to or smaller than the dimer energies have to be consider as part of the system in other approaches like NPI, and this always increases the simulation complexity. 

\begin{figure}
\includegraphics[width=8.5cm]{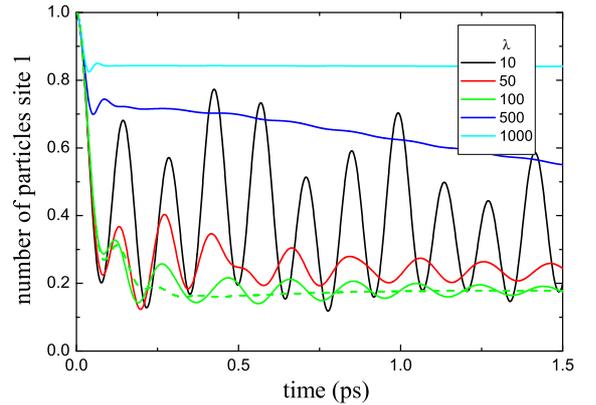}
\caption{Evolutions of the population on site $1$ for the spectral function of Eq. (\ref{JAR}) at various reorganization energies $\lambda$ and $T=0$ K. Dimer parameter are $J=100\textrm{cm}^{-1},\epsilon_{1}-\epsilon_{2}= 100\textrm{cm}^{-1}$.  Dashed line shows how the dynamics when the high-energy mode is decoupled. }
\label{fig4}
\end{figure}

The interaction with the near-resonant oscillator has a pronounced effect on the population dynamics, and Fig. \ref{fig4} shows how this coupling leads to a coherent beating effect which periodically suppresses population oscillations for $\lambda\leq 300 \mathrm{cm}^{-1}$. 
In situations where site $2$ might transfer population to another system, such a coherent suppression of oscillations could lead to an enhancement of EET from the dimer to that system. As $\lambda$ increases, the continuous part of the spectral density dominates the dynamics and we observe qualitatively similar behaviour to the dynamics obtained in Fig. \ref{fig3}. We note that the trapping-like dynamics for large $\lambda$ is less severe for this super-Ohmic $J(\omega)$, although the dynamics are still strongly non-Markovian for strong coupling.

A particularly striking feature of Fig. \ref{fig4} is that in the regime of optimal EET ($\lambda \sim 100 \mathrm{cm}^{-1}$), the high-energy mode leads to low amplitude oscillations which persist for at least $1.5$ ps. When the high-energy mode is decoupled, coherent oscillations vanish for $\lambda=100\mathrm{cm}^{-1}$ after just $~0.3$ ps. Experimental observation of such persistent undamped oscillations after a fast population transfer could thus indicate the presence of discrete high-energy modes in the environment of PPCs, and could be a useful signature for determining realistic $J(\omega)$s in PPCs.

{\em Discussion \& Conclusions --}
Combining an analytical chain transformation with numerically exact t-DMRG methods, we have introduced an efficient method
for the simulation of archetypal models of open quantum systems.
We have presented EET simulations of a dimer molecule in the presence
of a variety of environmental coupling structures and observed a rich
array of non-perturbative and non-markovian effects in the dynamics. Having demonstrated the viability of the technique, we plan to explore the important question of the role and persistence of coherence in
biological EET, using existing DMRG techniques to extend our method to both
finite temperatures, multiple sites, and spatially-correlated baths. As mentioned
previously, the method is general and we expect it to be relevant for a variety
of problems in condensed matter and quantum information, including fundamental
studies of decoherence, such as the spin-boson model, and the boundary of
quantum and classical physics.

\begin{acknowledgments}
This work is supported by the EU STREP projects CORNER, the EU Integrated Project QAP
and an Alexander von Humboldt Professorship. JP was supported by the Fundacion Seneca grants, 11602/EE2/09 and 11920/PI/09-J and Ministerio de Ciencia e Innovacion project number FIS2009-13483-C02-02. We thank A. Ishizaki and M. Thorwart for very useful discussions.
\end{acknowledgments}

\end{document}